\documentclass[a4paper,11pt]{article}
\usepackage{pos}
\usepackage{url}
\usepackage{multicol}
\usepackage{wrapfig}

\title{Toward N$^3$LO CT global PDF analyses: The SACOT-$\chi$ scheme in DIS and DGLAP evolution at partial N$^3$LO}
 \ShortTitle{The SACOT-$\chi$ scheme in DIS and DGLAP evolution at partial N$^3$LO}

\author*[a]{M. Guzzi}
\author[b]{Y. Fu}
\author[a]{C. Hampson}
\author[b]{P. Nadolsky}
\author[b]{M. Ponce-Chavez}
\author[b]{T.~Sharma}
\author[c]{K.~Xie}
\author[b]{C.-P. Yuan}

\affiliation[a]{Department of Physics, Kennesaw State University,
Kennesaw, GA 30144, USA}

\affiliation[b]{Department of Physics and Astronomy, Michigan State University,
East Lansing, MI 48824, USA}

\affiliation[c]{State Key Laboratory of Dark Matter Physics \& Key Laboratory for Particle Astrophysics
and Cosmology (MOE) \& Shanghai Key Laboratory for Particle Physics and Cosmology,
Tsung-Dao Lee Institute \& School of Physics and Astronomy, Shanghai Jiao Tong University, Shanghai 201210, China}

\emailAdd{mguzzi@kennesaw.edu}
\emailAdd{fuyao3@msu.edu}
\emailAdd{champso1@students.kennesaw.edu} 
\emailAdd{shar1157@msu.edu}
\emailAdd{nadolsky@msu.edu}
\emailAdd{xiekepi1@msu.edu}
\emailAdd{poncedan@msu.edu}
\emailAdd{yuanch@msu.edu}

\abstract{We discuss three aspects of the ongoing efforts of the CTEQ--TEA group toward global parton distribution function (PDF) analyses beyond NNLO accuracy in QCD. First, we present a realization of the SACOT-$\chi$ general-mass variable flavor number scheme for deep-inelastic scattering (DIS) at the partial next-to-next-to-next-to-leading order (pN$^3$LO) in QCD. We numerically illustrate how this SACOT-$\chi$ theory prediction captures the dependence of DIS structure functions on masses of heavy quarks with inclusion of the available ${\cal O}(\alpha_s^3)$ contributions. 
Second, we discuss a preliminary pN$^3$LO CT25 fit that implements such contributions.
Third, we extend the \texttt{Candia} algorithm  for solving DGLAP evolution equations for unpolarized parton distribution functions (PDFs) in $x$ space to N$^3$LO in QCD. The algorithm is based on logarithmic expansions of the solution and can be extended to all orders in QCD. An expansion equivalent to the exact DGLAP solution at N$^3$LO 
is presented in the non-singlet sector. We show results for approximate N$^3$LO PDFs evolved using the most recent approximations to the 4-loop splitting 
functions in QCD.}

\FullConference{The 33rd International Workshop on Deep Inelastic Scattering and Related Subjects (DIS2026)\\
4 - 8 May 2026\\
Bologna, Italy\\}

\begin{document}
\maketitle

{\bf Introduction.}
Deep-inelastic scattering (DIS), including high-precision inclusive and heavy-flavor DIS
measurements~\cite{H1:2015ubc,H1:2018flt} at HERA collider, provide the essential input to modern global analyses of parton distribution functions (PDFs) in the nucleon~\cite{Hou:2019efy,Bailey:2020ooq,NNPDF:2021njg,Ablat:2024muy}. DIS measurements constrain the flavor structure of PDFs over a broad range of Bjorken $x$
and momentum-transfer scales $Q$. Their interpretation requires control of heavy-quark (HQ) mass effects, including phase-space suppression near production thresholds and logarithmically enhanced radiative corrections at high $Q$ scales. Depending on the kinematics, the mass effects can be comparable in magnitude to the higher-order corrections targeted by NNLO and N$^3$LO analyses in QCD.

General-mass variable-flavor-number (GM-VFN) schemes provide a factorization framework for describing DIS cross sections with massive quark species across these kinematic regimes~\cite{Aivazis:1993kh,Aivazis:1993pi,
Buza:1996wv,Thorne:1997ga,Collins:1998rz,Tung:2001mv,Kramer:2000hn,Alekhin:2009ni,Forte:2010ta,Guzzi:2011ew,Stavreva:2012bs,Kusina:2013slm}.
They retain the relevant HQ mass dependence near threshold and resum large collinear logarithms through PDF evolution at high scales. The number of active parton flavors changes with the factorization scale, with matching relations connecting the corresponding PDFs and strong coupling. For $Q^2\gg m_h^2$, where $m_h$ denotes the heavy-quark mass, powers of $\ln(Q^2/m_h^2)$ can spoil the convergence of a fixed-order expansion and motivate this resummation. Consistent matching requires subtraction of the
overlap between the massive fixed-order contributions and the resummed contributions involving HQ PDFs. The resulting
cancellations must be maintained through the working perturbative order and implemented with sufficient numerical precision.

Extending global PDF analyses beyond NNLO, as in recent approximate N$^3$LO (aN$^3$LO) determinations~\cite{NNPDF:2024nan,McGowan:2022nag}, requires a consistent combination of higher-order parton evolution, hard-scattering coefficients, and HQ matching, together
with the corresponding extension of GM-VFN schemes. 
Recent public evolution codes~\cite{Bertone:2017gds,Candido:2022tld,Karlberg:2025hxk,Hampson:2025pvi,Hampson:2026yso} provide tools for incorporating the most up-to-date higher-order calculation for the DGLAP splitting functions, where the non-singlet (NS) sector is known exactly at N$^3$LO in QCD~\cite{Gehrmann:2026qbl,Moch:2026qsw} while the singlet (S) sector uses approximations ~\cite{Davies:2022ofz,Moch:2021qrk,Falcioni:2023luc,Falcioni:2023vqq,Moch:2023tdj,Falcioni:2024xyt,Falcioni:2024qpd,Falcioni:2025hfz}.
Outside of global fits, N$^3$LO hard-scattering calculations are available for selected processes and observables, including Higgs-boson production in specified theoretical approximations and Drell--Yan production~\cite{Anastasiou:2015vya,Mistlberger:2018etf,Dreyer:2016oyx,Duhr:2020sdp,Duhr:2021vwj}. In neutral-current DIS discussed here, massless coefficient functions~\cite{Vermaseren:2005qc,Moch:2004xu,Moch:2007rq,Moch:2008fj} as well as massless single-jet production contributions~\cite{Gehrmann:2018odt,Currie:2018fgr} are known to $\mathcal{O}(\alpha_s^3)$.
For heavy-flavor DIS, however, the Wilson coefficients at $\mathcal{O}(\alpha_s^3)$ with full mass
dependence are only known in several asymptotic kinematical regimes. Recent calculations have completed the three-loop single-mass heavy-flavor operator
matrix elements (OMEs)~\cite{Behring:2025avs,Ablinger:2025nnq,Ablinger:2025joi,Ablinger:2024qxg,Ablinger:2024xtt},
enabling the determination of the corresponding Wilson coefficients in the asymptotic region $Q^2\gg m_h^2$~\cite{Ablinger:2025awb}.
These results do not determine the power-suppressed mass contributions relevant away from this limit. Descriptions covering the full kinematic range therefore require approximations to the missing massive contributions~\cite{Kawamura:2012cr,
Barontini:2026drp,Laurenti:2021akt,Alekhin:2017kpj}. The associated theoretical uncertainties must be assessed when incorporating these contributions into higher-order global PDF analyses. 

{\bf Neutral-current heavy-flavor DIS in S-ACOT-$\chi$ at ${\cal O}(\alpha_s^3)$. \label{sec:acot}}
At the DIS'2026 workshop, we summarized the ongoing implementation of the N$^3$LO contributions in the CTEQ-TEA global analysis and the first CT25 global fit based on the pN$^3$LO. Following the CTEQ-TEA implementations at NNLO, we work in the SACOT-$\chi$ GM-VFN scheme \cite{Guzzi:2011ew}, which is a DIS realization of a more comprehensive "SACOT with Massive Phase Space" (SACOT-MPS) scheme applicable both in lepton-hadron and hadron-hadron processes \cite{Xie:2021ycd,Guzzi:2024can}. 

QCD factorization~\cite{Collins:1998rz} expresses an inclusive DIS structure function as a convolution,
\begin{equation}
 F_A(x,Q)=\sum_i e_i^2\sum_a\left[C_{i,a}\otimes \Phi_{a/A}\right](x,Q), \mbox{ with }
  [C\otimes f](x)=\int_x^1\frac{d\xi}{\xi}\,C\!\left(\frac{x}{\xi}\right)f(\xi).
 \label{eq:factorization}
\end{equation}
Here the notations follow the SACOT-$\chi$ NNLO implementation in Ref.~\cite{Guzzi:2011ew}.
The index $i$ 
runs over all quark flavors $i = 1,\dots, N_f^{fs}$ that couple to the virtual photon with fractional electric charges $e_i = 2/3$ or $-1/3$. The index $a$ represents parton flavors (quark flavors and the gluon) in the PDF $\Phi_{a/A}(\xi,\mu)$ in the parent hadron $A$.
The coefficient functions $C_{i,a}$ depend on $Q/\mu$, $m_h/\mu$, and the strong coupling $\alpha_s(\mu)$, with $Q$ and $\mu$ denoting the momentum transfer and factorization scale. 

Given the structure function $F_b(x,Q)\equiv \sum_i e_i^2 F_{i,b}(x,Q)$ for DIS $e+b\to e+X$ on an initial-state parton $b$, the $\mathcal{O}(\alpha_s^n)$ coefficient functions $C_{i,b}^{(n)}$ are obtained from $F_{i,b}^{(n)}$ by removing collinear contributions already contained in the PDFs~\cite{Guzzi:2011ew,Guzzi:2024can}. In terms of the parton-in-parton distributions given by the OMEs $A_{ab}^{(n)}$, the resulting subtraction pattern for the $\mathcal{O}(\alpha_s^3)$ coefficients is
\begin{align}
 C_{i,b}^{(3)}={}&F_{i,b}^{(3)}
 -C_{i,a}^{(0)}\otimes A_{ab}^{(3)}
 -C_{i,a}^{(1)}\otimes A_{ab}^{(2)}
 -C_{i,a}^{(2)}\otimes A_{ab}^{(1)}\,.
 \label{eq:acot-subtraction}
\end{align}
For a massive quark, the OMEs contain the logarithmic dependence on the matching scale,
$A_{Qj}^{(n)}=\sum_{\ell=0}^{n}\ln^{\ell}(\mu^2/m_h^2)\,a_{Qj}^{(n,\ell)}$. This sum can be reorganized in a number of ways. For instance, when computing the inclusive structure function $F_2(x,Q)$ in the effective QCD theory with one massive quark flavor denoted by $h$, massless quarks denoted by $l$, and the gluon $g$, the $\mathcal{O}(\alpha_s^3)$ component $F^{(3)}_{2,h}(x,Q)$ in which the photon couples to the heavy quark $h$ is given by 
\begin{equation}
\begin{aligned}
a_s^3 F^{(3)}_{2,h}
=&a_s^3 \left(\widehat{F}_{h,g}^{(3)}-c_{h,h}^{(0)}\otimes A_{hg}^{(3)}
-c_{h,h}^{(1)}\otimes A_{hg}^{(2)}
-c_{h,h}^{(2)}\otimes A_{hg}^{(1)}
\right)\otimes f_g
\\
+&a_s^3 \left(\widehat{F}_{h,l}^{(3)}
-c_{h,h}^{(0)}\otimes A_{hl}^{(3)}
-c_{h,h}^{\rm NS (1)}\otimes A_{hl}^{(2)}
\right)\otimes f_\Sigma
+a_s^2 c_{h,h}^{(2)}\otimes f_{h},
\end{aligned}
\label{eq:f3}
\end{equation}
where $a_s\equiv \alpha_s/4\pi$, $f_\Sigma$ and $f_h$ are the light-quark singlet and heavy-quark PDFs, and the caret in $\widehat F_{...}$ indicates that the massless collinear poles were subtracted.

In Eq.~(\ref{eq:f3}), 
$\widehat F_{h,g}^{(3)}$, the dominant $\mathcal{O}(\alpha_s^3)$ piece near the threshold, has been calculated only in several kinematic limits. Its behavior over the entire kinematic domain strongly depends on the choice of interpolation~\cite{Barontini:2026drp}.
To estimate $\widehat F_{h,g}^{(3)}$,  the current CTEQ--TEA approximation follows the strategy of Refs.~\cite{Kawamura:2012cr,
Barontini:2026drp,Laurenti:2021akt,Alekhin:2017kpj}: information from the threshold, high-energy (small-$x$), and high-scale $Q^2\gg m_h^2$ limits is "glued together" using smooth damping functions.  In our calculation, the high-energy prediction has been updated by implementing the latest three-loop OMEs and massive Wilson coefficients~\cite{AblingerVFNS2026,AblingerF22026}. 

Equation~(\ref{eq:f3}) contains distinct dynamical components on the right-hand side: positive flavor-creation (FC) terms convolved with $f_g$ and $f_\Sigma$, flavor-excitation (FC) terms convolved with $f_h$, and the negative subtraction (SUB) pieces that cancel the dominant parts of either FE or FC respectively in the $Q^2 \approx m_h^2$ and $Q^2\gg m_h^2$ limits. To achieve smooth interpolation over the whole $Q$ domain, this cancellation must be realized with consistent perturbative accuracy and high numerical precision.

\begin{figure}[tb]
\centering
\begin{minipage}[t]{0.49\textwidth}
\includegraphics[width=1\linewidth]{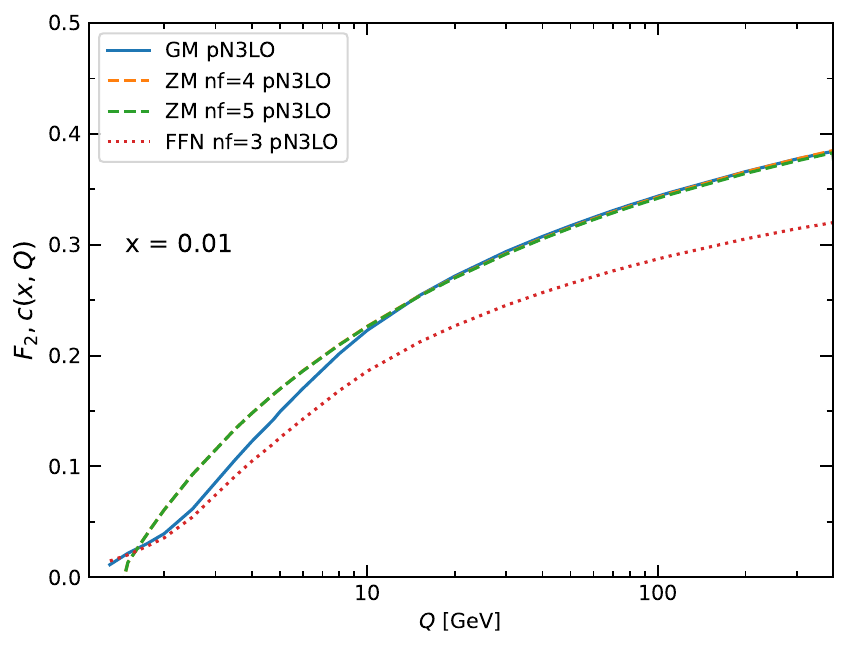}
\caption{Theory predictions for the charm component $F_{2,c}$ of the inclusive structure function $F_2(x,Q)$ at partial N$^3$LO (pN$^3$LO) in the SACOT-$\chi$ (GM), ZM ($N_f=4$ and $5$), and FFN ($N_f=3$) schemes. The input CT25NNLO PDFs at $Q_0$ are evolved at N$^3$LO with \texttt{HOPPET-v2} \cite{Karlberg:2025hxk}. The GM result agrees very well with the FFN and ZM results at $Q\approx m_c=1.3$ GeV and $Q \approx 100$ GeV, respectively. Similar quality of matching is achieved at other $x$ values.}
\label{fig:f2c}
\end{minipage}
\hfill
\begin{minipage}[t]{0.49\textwidth}
\centering
\includegraphics[width=1\linewidth]
{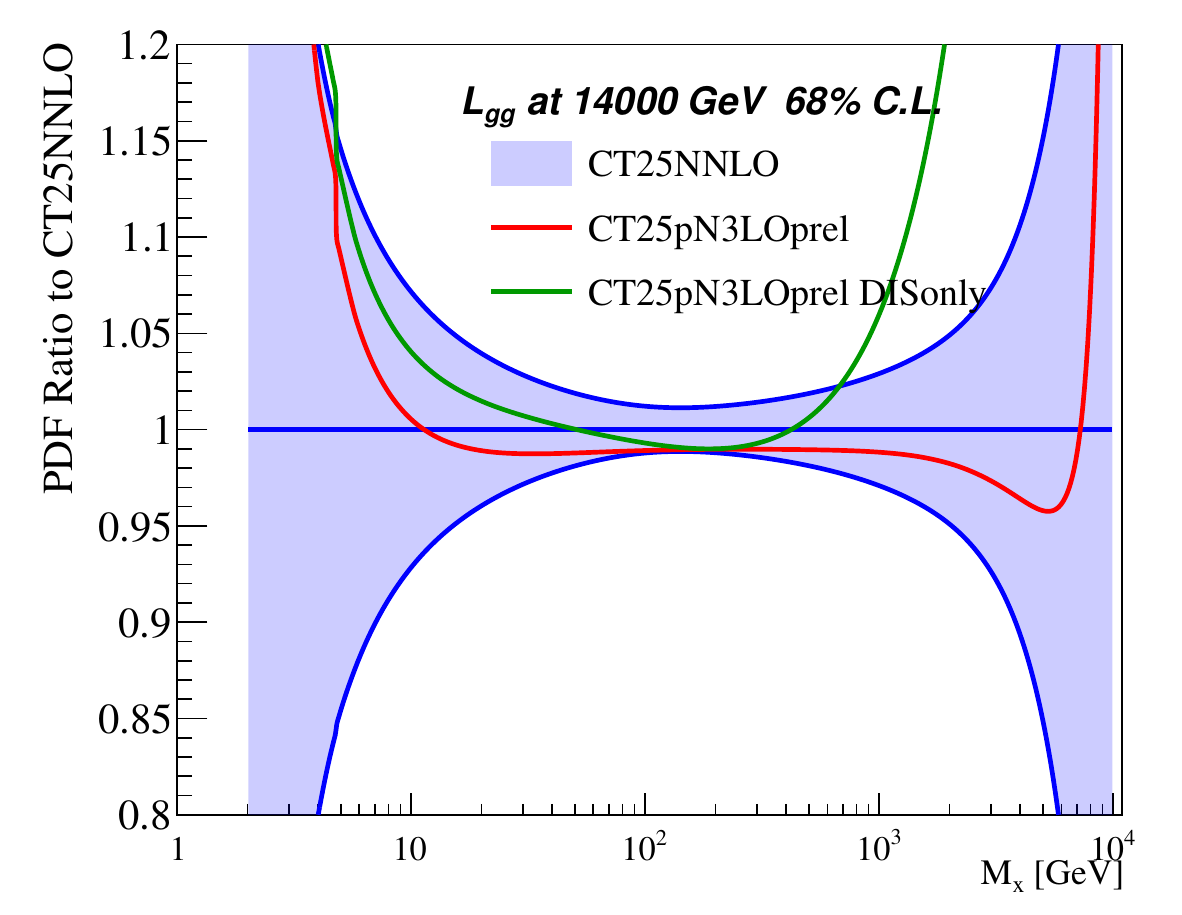}
\caption{Preliminary gluon--gluon luminosity comparison at
$\sqrt{s}=14$~TeV, normalized to CT25 NNLO. The shaded band represents the CT25 NNLO 68\% confidence-level PDF uncertainty. The red and green curves show the preliminary pN$^3$LO variants, labeled CT25pN3LOprel and CT25pN3LOprel DISonly, respectively. Near the Higgs mass, the pN$^3$LO DIS treatment lowers the luminosity by approximately $1.2\%$ relative to CT25NNLO. }
\label{fig:ct25-gg-luminosity}
\end{minipage}
\end{figure}

Figure~\ref{fig:f2c}, showing the S-ACOT-$\chi$ result up to pN$^3$LO for the charm component $F_{2,c}$ of the inclusive $F_2$ at $x=0.01$, demonstrates that we achieved such stable behavior. The figure shows how 
the pN$^3$LO SACOT-$\chi$ (GM) result interpolates between
the mass-dependent fixed-flavor number (FFN $N_f=3$) and zero-Mass (ZM $N_f=5$) predictions. To achieve this interpolation with a percent-level accuracy, the SACOT-$\chi$ implements the complete
$N_f\rightarrow N_f+1$ scheme transformations~\cite{Bierenbaum:2009zt}, including the resulting logarithmic $\ln^p(Q^2/\mu^2)$ contributions that are important for N3LO benchmarking of $F_{2,c}$. 

{\bf Subtraction and Residual PDFs.} A useful reorganization of the ACOT cancellation patterns that can improve numerical accuracy is obtained by collecting together the subtraction (SUB) and FE terms (which share the same matrix elements) by introducing a universal subtraction PDF~\cite{Guzzi:2024can}: 
\begin{align}
\mathrm{SUB}
&= c_{hh}^{(0)} \otimes
  \left(
    a_s A_{hg}^{(1)}
    + a_s^2 A_{hg}^{(2)}
    + a_s^3 A_{hg}^{(3)}
  \right) \otimes g
+ c_{hh}^{(0)} \otimes
  \left(
    a_s^2 A_{hl}^{(2),\mathrm{PS}}
    + a_s^3 A_{hl}^{(3),\mathrm{PS}}
  \right) \otimes \Sigma
+ \cdots ,
\\
\mathrm{FE}
&= c_{hh}^{(0)} \otimes
  \left(
    a_s f_h^{(1)}
    + a_s^2 f_h^{(2)}
    + a_s^3 f_h^{(3)}
    + \cdots
  \right)
= c_{hh}^{(0)} \otimes f_h + \cdots ,
\\
\widetilde f_h^{\rm N^3LO}
&=\left(
    a_s A_{hg}^{(1)}
    + a_s^2 A_{hg}^{(2)}
    + a_s^3 A_{hg}^{(3)}
  \right) \otimes g
+ c_{hh}^{(0)} \otimes
  \left(
    a_s^2 A_{hl}^{(2),\mathrm{PS}}
    + a_s^3 A_{hl}^{(3),\mathrm{PS}}
  \right) \otimes \Sigma
+ \cdots ,  
\end{align}
where $\widetilde f_h^{\rm N^3LO}$ represents the subtraction PDF defined as the HQ distribution generated perturbatively from the OME expansion through ${\cal O}(\alpha_s^3)$. The residual PDF is defined as the difference
\begin{align}
\delta f_h^{\rm N^3LO}=f_h-\widetilde f_h^{\rm N^3LO},
\label{eq:residual}
\end{align}
such that FE and SUB terms can be combined schematically as
\begin{equation}
 F_{2,h}\simeq F_{2,h}^{\rm FC}
 +c_{h,h}^{(0)}\otimes\delta f_h^{\rm N^3LO}+\ldots .
 \label{eq:residual-structure}
\end{equation}
This representation directly shows the residual left after the perturbatively generated part of the HQ PDF cancels the subtraction term. Subtraction- and Residual-PDF grids have been constructed in LHAPDF format~\cite{subPDFgrids,sacotmps} for public use. The pN$^3$LO DIS structure functions have also been incorporated into the C++ \texttt{DragonWell} framework used in ongoing CTEQ-TEA studies, which will be made publicly available.

{\bf Preliminary pN$^3$LO impact on the gluon--gluon luminosity.\label{sec:gg-luminosity}}
We combined the described pN$^3$LO implementation for the heavy-quark NC DIS component with the full N$^3$LO coefficient functions for light flavors and then performed a preliminary CT25pN$^3$LO fit to access the impact on the PDFs. Figure~\ref{fig:ct25-gg-luminosity} compares the CT25 NNLO gluon--gluon luminosity, $L_{gg}(M_X)$ at $\sqrt{s}=14$~TeV, to the ones from two such fits: CT25pN3LOprel, which computes NC DIS at pN$^3$LO, while fitting the rest of the CT25 data sets at NNLO; and CT25pN3LOprel-DISonly, which fits the CT25 parametrizations to DIS data sets only. \texttt{HOPPET-v2}~\cite{Karlberg:2025hxk} is used for parton evolution. Relatively to CT18NNLO~\cite{Hou:2019efy}, the cumulative difference in $L_{gg}(M_X)$ in the Higgs mass region, $M_X=M_H=125$ GeV, reflects several effects. The new LHC datasets in the CT25NNLO baseline lower the luminosity near $M_H$ by about $2.5\%$. Inclusion of pN$^3$LO terms in the NC DIS further reduces $L_{gg}(M_H)$ by approximately $1.2\%$, produces an overall decrease of roughly $4\%$ with respect to CT18NNLO. The corresponding description of the 1120 HERA DIS points changes
only slightly: $\chi^2$ decreases from 1401 to 1399, or from 1.251 to 1.250 per point. The luminosity shift therefore accompanies a modest change in the overall quality of the fit to HERA data. While these results are still preliminary, the reduction in the CT25pN3LOprel gluon in the Higgs region is consistent
with the MSHT and NNPDF aN3LO PDFs. However, the uncertainties associated with incomplete N$^3$LO ingredients are potentially sizable. The PDF uncertainty band alone does not quantify them.

{\bf N$^3$LO DGLAP evolution with \texttt{Candia-v2}.\label{sec:candia}}
Besides the advancements in the DIS structure functions calculation, the other key ingredient in N$^3$LO global PDF analyses is parton evolution. In this section, we summarize the recent extension to N$^3$LO in QCD of the $x$-space \texttt{Candia} algorithm for solving DGLAP evolution.
This algorithm uses the recent progress in the splitting function and OME calculations to solve the DGLAP equations directly in $x$ space through recursion relations obtained from logarithmic expansions~\cite{Cafarella:2008du,Guzzi:2006wx,Cafarella:2005zj},  
which produce exact solutions in power series in the NS sector. The new C++ implementation, \texttt{Candia-v2}, extends this construction to N$^3$LO~\cite{Hampson:2025pvi,Hampson:2026yso}.  
The $m$-order exponential-type power series ansatz in $x$-space at the generic scale $\mu$ for the NS sector, and the $\kappa$-truncated logarithmic ansatz for both NS and S sectors, are given as
\begin{align}
f^{\mathrm{N}^m\mathrm{LO}}(x,\mu^2) =
\prod_{i=0}^m \left[
\left(\sum_{n=0}^\infty \frac{a_{n,i}(x)}{n!}L_i^n\right)_{\otimes}
\;
\right]
f(x,\mu_0^2), 
~~~
{\bf f}(x,\mu^2)=\sum_{i=0}^{\kappa}
\left[\sum_{n=0}^{\infty}\alpha_s^i {\bf A}_{n,i}(x)
\ln^n\left(\frac{\alpha_s(\mu^2)}{\alpha_s(\mu_0^2)}\right)\right] ,
\label{eq:candia-ansatz}
\end{align}
where $m$ corresponds to the expansion order of the splitting functions, and $\kappa$ represents the truncation index that defines the $\alpha_s$ accuracy of solution in both NS and S sectors\footnote{Note that in Eq.~(\ref{eq:candia-ansatz}) the $\kappa$-truncated ansatz uses boldface characters for $f$ and $A_{n,i}$ to indicate vector solutions in the singlet sector case. The $L_i^n$ represent logarithmic functions of $\alpha_s(\mu)$, $\alpha_s(\mu_0)$, and $\beta_i$ in the NS case.}. $\kappa$ can formally be larger than $m$ to obtain higher-order truncated solutions. Substitution of Eqs.~(\ref{eq:candia-ansatz}) into the evolution equation determines the coefficients $a^i_n(x)$ and ${\bf A}_n^{i}(x)$ recursively.  
At NLO, NNLO and N$^3$LO the recursion relations contain the $P^{(i)}(x)$ splitting functions  and $\beta_{i}$ terms with $i=0,1,2,3$. The resulting truncated logarithmic solutions are equivalent to the familiar $U$-matrix construction~\cite{Cafarella:2008du,Guzzi:2006wx,Cafarella:2005zj}. An important distinction is that the evolution \emph{algorithm} is exact at the stated perturbative order for the kernels supplied to it.  The label pN$^3$LO for the evolved PDFs reflects the fact that parts of the four-loop splitting-function matrix in the S sector are available only through high-accuracy approximations. 
\begin{wrapfigure}{r}{0.71\textwidth}
\centering
\includegraphics[width=1\linewidth]{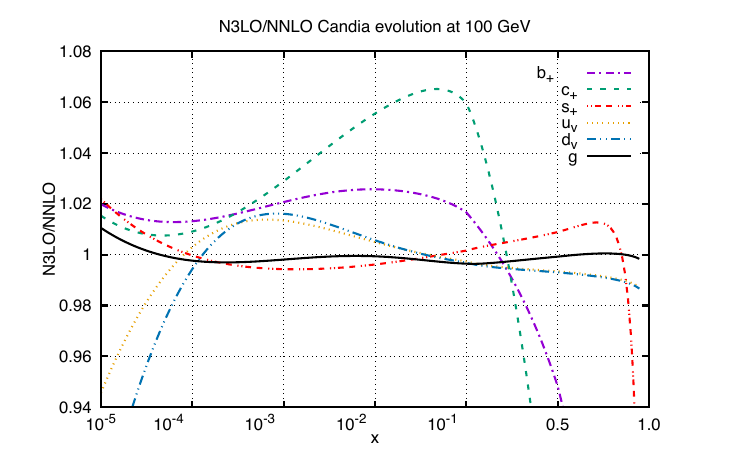}
\caption{N$^3$LO/NNLO PDF ratios in the VFNS scheme at $\mu_F=100$ GeV with N$^3$LO matching conditions. Figure from ref.~\cite{Hampson:2025pvi}.}
\label{fig:candia}
\end{wrapfigure}
Figure~\ref{fig:candia} shows the impact of N$^3$LO evolution using a N$^3$LO/NNLO ratio plot as a function of $x$ at $\mu=100$ GeV for selected PDF flavors. The Les Houches initial conditions of Ref.~\cite{CooperSarkar2024} are used to facilitate benchmarking.
A benchmark with \texttt{HOPPET-v2}~\cite{Karlberg:2025hxk} is performed in~\cite{Hampson:2025pvi}, where agreement is found at the permille level over the tabulated grid, with small deviations appearing toward kinematic endpoints.  
The rewrite from the original \texttt{C} implementation also improves the numerical organization and optimizes memory usage~\cite{Hampson:2026yso,Hampson:2025pvi}. For example, the concurrent evolution of independent NS combinations is now possible, together with the parallelization of expensive singlet convolutions.

{\bf Summary and outlook.\label{Conclusions}}
This contribution summarizes, on the one hand, recent developments by the CTEQ--TEA group on the extension of the SACOT-$\chi$ scheme for neutral-current DIS to pN$^3$LO in QCD, phenomenological predictions for massive-quark DIS structure functions, and preliminary CT25 PDF fits at pN$^3$LO. 
The latest SACOT-$\chi$ implementation provides pN$^3$LO heavy-flavor structure functions using the known three-loop matching information and the best presently available approximation to the full-mass ${\cal O}(\alpha_s^3)$ coefficient functions.  The subtraction/residual-PDF formulation makes the FC--SUB--FE cancellation transparent and is being used within the C++ fitting framework \texttt{DragonWell}.  
On the other hand, we discussed the recent extension of the \texttt{Candia} algorithm for solving DGLAP evolution in $x$ space to N$^3$LO in QCD. DGLAP solutions at pN$^3$LO are obtained through recursion relations based on a logarithmic ansatz that leads to an exact solution as a power series for the non-singlet sector. The algorithm is implemented in the publicly available \texttt{C++} code \texttt{Candia-v2}, which is in agreement at the permille level with \texttt{HOPPET-v2}. These developments provide complementary building blocks for forthcoming CTEQ--TEA PDF studies at N$^3$LO.

{\bf Acknowledgments.} We thank Johannes Bl\"umlein for discussions and correspondence. The work of MG and CH was supported by the U.S. National Science Foundation (NSF) under Grant No. PHY-2412071. The work of KX was supported in part by the U.S. NSF under Grant No. PHY-2310497. The work at Michigan State University was partly supported by the U.S. Department of Energy, Office of Science, Office of High Energy Physics under Award DE-SC0026593,  by the U.S. NSF under Grant No. PHY-2310291, and by the Wu-Ki Tung Endowed Chair in particle physics.  This work was supported in part through computational resources and services provided by the Institute for Cyber-Enabled Research at Michigan State University, and used the high-performance computing resource from the Kennesaw State University HPC clusters.

\begingroup
\renewcommand{\bibfont}{\fontsize{9}{10}\selectfont}
\setlength{\bibsep}{0pt}
\setlength{\columnsep}{14pt}
\begin{multicols}{2}[\section*{References}]
\renewcommand{\bibsection}{}
\bibliographystyle{compact}
\bibliography{references}
\end{multicols}
\endgroup

\end{document}